\documentclass[prx,twocolumn,superscriptaddress,amsmath,amssymb,longbibliography]{revtex4-1}

\usepackage{graphicx}
\usepackage{color}
\usepackage{ulem}

\newcommand{\eqreff}[1]{Eq.~(\ref{#1})}
\newcommand{\Eqreff}[1]{Eq.~(\ref{#1})}

\newcommand{\be}{\begin{equation}}
\newcommand{\ee}{\end{equation}}
\newcommand{\bes}{\begin{eqnarray}}
\newcommand{\ees}{\end{eqnarray}}
\newcommand{\bess}{\begin{eqnarray*}}
\newcommand{\eess}{\end{eqnarray*}}

\begin{document}

\title{Roles of Local Non-equilibrium Free Energy in the Description of Biomolecules
}

\author{Lee Jinwoo}
\affiliation{Department of Mathematics, Kwangwoon 
University, 20 Kwangwoon-ro, Nowon-gu, Seoul 01897, Korea}
\email{jinwoolee@kw.ac.kr}

\date{\today}

\begin{abstract}
When a system is in equilibrium, external perturbations yield a time series of non-equilibrium distributions, and recent experimental techniques give access to the non-equilibrium data that may contain critical information. Jinwoo and Tanaka (L. Jinwoo and H. Tanaka, Sci. Rep. 2015, 5, 7832) have provided mathematical proof that such a process's non-equilibrium free energy profile over a system's substates has Jarzynski's work as content, which spontaneously dissipates while molecules perform their tasks. Here we numerically verify this fact and give a practical example where we analyze a computer simulation of RNA translocation by a ring-shaped ATPase motor. By interpreting the cyclic process of substrate translocation as a series of quenching, relaxation, and second quenching, the theory gives how much individual sub-states of the ATPase motor have been energized until the end of the process. It turns out that the efficiency of RNA translocation is $48\sim 60\%$ for most molecules, but $12\%$ of molecules achieve $80\sim 100\%$ efficiency, which is consistent with the literature. This theory would be a valuable tool for extracting quantitative information about molecular non-equilibrium behavior from experimental observations.  
\end{abstract}

\maketitle
\section{Introduction}
Counting is the basis of thermodynamics, which describes the behavior of biomolecules~\cite{callen}. Free energy is (negative logarithm of) the number of configurations with more weights to lower energy conformations~\cite{jarReview}. If one restricts to a specific sub-state, it gives the concept of conformational free energy, (negative logarithm of) the weighted number of conformations that belong to that sub-state. Thus when the second law of thermodynamics states that molecules tend to minimize their free energy, it means that molecules would be in (sub)states that include more accessible configurations with lower energies.

We may transform a system's (conformational) free energy by supplying energy through external control, such as in the case of single-molecule experiments~\cite{liph2001,liph2002, expColin,ritort2012}. Since molecules behave stochastically, we build an ensemble by repeating experiments to apply thermodynamics in understanding molecules' behavior~\cite{revSears2008, jarReview, revSeifert}. For example, Jarzynski's work fluctuation theorem enables one to convert fluctuating work into the difference of free energies between the end states of external control~\cite{jar}. 

On the one hand, Jarzynski's work fluctuation theorem can be very informative since free energy provides comprehensive information about the state of a system~\cite{hummer2010,toyabe2010,multiple2010,ritort2012}. On the other hand, it might be insufficient to provide detailed information on molecular non-equilibrium behavior on the level of sub-states~\cite{boyer1997atp,liepelt2007,motor2007}. For example, when a molecular machine hydrolyzes ATP, hydrolysis energy will probably increase the molecules' energy, allowing it to overcome the free energy barrier. However, since the process proceeds in an entire non-equilibrium situation and free energy provides only macro-information, a detailed description of this process is elusive.

Jinwoo and Tanaka have recently revealed that those trajectories that reach the individual sub-state of a system contain essential details for the thermodynamics of each sub-state~\cite{local}. Significantly, they show that each substate's local non-equilibrium free energy has Jarzynski's work as content, allowing us to see how the introduced energy by external control affects molecular sub-states. 

In this paper, we numerically verify Jinwoo and Tanaka's local version of Jarzynski's work fluctuation theorem. Mainly, we elucidate local non-equilibrium free energy, which contains detailed information about biomolecules performing their functions while allocating energy between sub-states and dissipating that energy. As an application, we analyze a simulation carried by Ma and Schulten for RNA translocation by a ring-shaped ATPase motor~\cite{ma2015mechanism}. There, we reveal Jarzynski's work required for reaching each substate at the end of the translocation process. 


We organize the paper as follows: Section 2 introduces the theoretical framework while clarifying the terms used. In Section 3, firstly, we provide simulation results, confirming that local non-equilibrium free energy and an ensemble average of Jarzynski's work coincide at each time and substate. Secondly, we provide an application showing that the theory enables one to extract critical quantities from experiments or simulations. Section 4 discusses the implication of the results. 

\section{Theoretical Framework}
\subsection{Terminologies}
Recent non-equilibrium theories~\cite{fisher1999, fisher2001, bustamante2001, schmiedl2008, hwang2016,hwang2018,sasa,expSasa,seifert05,sagawa2,jinwoo2019fluctuation,jinwoo_2019} are  based on the following terminologies of equilibrium thermodynamics.
Let us consider a molecular system in the heat bath of inverse temperature $\beta=1/(k_BT)$, where $k_B$ is the Boltzmann constant, and $T$ is the heat bath temperature. External control $\lambda_t$ at time $t$ of the system determines the system's free energy $F(\lambda_t)$, (negative logarithm of) the number of configurations with more weights to lower energy configurations:
\begin{equation}\label{eq:F}
F(\lambda_t) = -\frac{1}{\beta}\ln\left[\int_{\Omega(\lambda_t)} e^{-\beta E(z;\lambda_t)}\,dz\right],
\end{equation}
where $E(z;\lambda_t)$ is the internal energy of the system configuration $z$. \Eqreff{eq:F} indicates that the lower the free energy, the richer the accessible configurations with lower energies. We note that external control $\lambda_t$ directly changes the internal energy $E$ and the set of accessible configurations, $\Omega$.

If one considers free energy restricted to a sub-state $\Gamma$, it gives the concept of conformational free energy $G(\Gamma)$, (negative logarithm of) the weighted number of configurations restricted to that sub-state: 
\be\label{eq:G}
G(\Gamma;\lambda_t)=-\frac{1}{\beta}\ln\left[\int_{z\in \Gamma} e^{-\beta E(z;\lambda_t)}\,dz\right].
\ee
\Eqreff{eq:G} tells that the lower the conformational free energy of sub-state $\Gamma$, the richer the accessible microstates with lower energies. Thus with $\lambda_t$ fixed, after exploring the conformational free energy landscape for an extended period, the smaller the conformational free energy of sub-state $\Gamma$, the higher the probability that a system stays in $\Gamma$:
\be\label{eq:p_inf}
p_{\infty}(\Gamma;\lambda_t)  = \frac{e^{-\beta G(\Gamma;\lambda_t)}}{e^{-\beta F(\lambda_t)}}.
\ee

We consider a situation where we pre-determinedly varying external control $\lambda_t$ as time $t$ varies from $0$ to $\tau$.
A system's microstate $z_t$ at time $t$ would form a trajectory, $\{z_t\}_{0\le t\le\tau}$. 
If one accumulates along trajectory $\{z_t\}_{0\le t\le\tau}$ the increment of internal energy $E$ at each $z_t$ due to external control $\lambda_t$ defines 
Jarzynski's work:
\begin{equation}\label{eq:W}
W = \int_{\lambda_0}^{\lambda_\tau} \frac{\partial E(z_t;\lambda_t)}{\partial \lambda_t}\,d\lambda_t.
\end{equation}
On the other hand, if one accumulates along trajectory $\{z_t\}_{0\le t\le\tau}$ the increment of internal energy at each $z_t$ due to thermal fluctuation defines heat:
\be\label{eq:Q}
Q = \int_{z_0}^{z_\tau} \frac{\partial E(z_t;\lambda_t)}{\partial z_t}\circ dz_t,
\ee 
where $\circ$ indicates 'stochastic multiplication' in the Stratonovich sense. Combining $\eqreff{eq:W}$ and $\eqreff{eq:Q}$ gives the first law of thermodynamics along trajectory $\{z_t\}_{0\le t\le\tau}$~\cite{sekimoto}:  
\be\label{eq:1st-law}
\Delta E =  W + Q,
\ee
where $\Delta E = E(z_\tau;\lambda_\tau)-E(z_0;\lambda_0)$ indicates the increment of internal energy along the path.


\subsection{Microscopic Reversibility}
Since microscopic reversibility forms the basis of work fluctuation theorems, we discuss it briefly~\cite{kur, maes1999,jar2000}. 
To this end, we consider the time-reversed process the same as filming the forward process and playing it backward. 
For control $\lambda_t$ that varies from time $0$ to $\tau$, let time-reversed control $\lambda'_t\equiv \lambda_{\tau-t}$, where $\equiv$ denotes 'defined by'. As time $t$ goes from $0$ to $\tau$, time-reversed control $\lambda'_t$ strictly follows $\lambda_t$ in a time-reversed manner. 
For trajectory $\{z_t\}_{0\le t\le \tau}$, let time-reversed trajectory $\{z'_t\}_{0\le t\le\tau}\equiv\{z_{\tau-t}\}_{0\le t\le \tau}$, which again follows the forward trajectory in a time-reversed manner as time $t$ goes from $0$ to $\tau$.

The microscopic reversibility condition says that a lot of spontaneous heat flow from the heat bath to the system along a trajectory is less probable or a rare event. In detail, the conditional probability of time-forward path $\{z_t\}_{0\le t\le \tau}$ conditioned on the initial configuration $z_0$
decreases exponentially as heat flow from the heat bath to the system along the path increases~\cite{review}:
\be\label{eq:rev}
\frac{p\left[\{z_t\}_{0\le t\le \tau}|z_0\right]}{p'\left[\{z'_t\}_{0\le t\le\tau}|z'_0\right]}=e^{-\beta Q}.
\ee
Here the forward probability is compared to the conditional probability of the time-reversed trajectory $\{z'_t\}_{0\le t\le\tau}$ conditioned on the initial conformation $z'_0$ for the backward process. This comparison to the time-reversed probability $p'$ makes the condition \eqreff{eq:rev} invariant under time reversal.  
If one reversed time, the left-hand side of \eqreff{eq:rev} would flip, and heat $Q$ absorbed by the system from the heat bath would be released upon time reversal, flipping the right-hand side of \eqreff{eq:rev}, too. Thus, condition \eqreff{eq:rev} states precisely the same for the time-reversed trajectory, so its name derives.

\subsection{Work Fluctuation Theorems} 

Let us consider a single molecule (e.g., RNA) pulling experiment using an optical tweezer.
We assume the system is in equilibrium at external control $\lambda_0$~\cite{ritort2012,expSasa,ExpTest2018,liph2002}.  
As external control $\lambda_t$ pulls RNA during $0\le t\le \tau$, it would give energy to RNA by $W$ as in \eqreff{eq:W}, changing the free energy landscape of the molecule (see \eqreff{eq:F} and \eqreff{eq:G}, which depend on energy $E$). Repeating the experiment (which would hypothetically yield a non-equilibrium probability of sub-states $\Gamma$ at each time $t$) and taking the following average of Jarzynski's work $W$ would give the free energy difference between the two end states $\lambda_0$ and $\lambda_\tau$~\cite{jar2007work,jar2000}:
\be\label{eq:jar}
\left<e^{-\beta W}\right> = e^{-\beta \left[F(\lambda_\tau)-F(\lambda_0)\right]},
\ee
where $\left<\cdot\right>$ indicates the average over the repeated experiments, and $F$ is free energy as defined in \eqreff{eq:F}.

Jarzynski's work fluctuation theorem, \eqreff{eq:jar}, takes into account all the paths generated while the forward process repeats. On the other hand,
Jinwoo and Tanaka considers only those paths that reach a sub-state $\Gamma$ at final time $\tau$. In detail, they have shown that the following average of Jarzynski's work of paths that reach $\Gamma$ at time $\tau$ gives local non-equilibrium free energy $\mathcal{F}$ of sub-state $\Gamma$ as follows~\cite{local}:
\be\label{eq:JT}
\left<e^{-\beta W}\right>_{\Gamma,\tau} = e^{-\beta \left[\mathcal{F}(\Gamma,\tau)-F(\lambda_0)\right]},
 \ee
 where $\left<\cdot\right>_{\Gamma,\tau}$ on the left-hand side indicates the average over the conditioned paths that reach $\Gamma$ at time $\tau$. On the right-hand side of \eqreff{eq:JT} appears the difference between the local non-equilibrium free energy $\mathcal{F}$ of sub-state $\Gamma$ at time $\tau$ and free energy $F$ at external control $\lambda_0$. One may interpret \eqreff{eq:JT} as that the local non-equilibrium free energy of each sub-state at time $\tau$ has the ensemble of Jarzynski's work of each path that reaches that sub-state at time $\tau$ as content. Here, the local non-equilibrium free energy $\mathcal{F}$ of $\Gamma$ at $\tau$ is composed of conformational free energy $G$ of sub-state $\Gamma$ as in \eqreff{eq:G} and stochastic entropy, $-k_B\ln p$, as follows:
 \be\label{eq:calF}
 \mathcal{F}(\Gamma,t) = G(\Gamma;\lambda_t) + \frac{1}{\beta}\ln p(\Gamma,t),
 \ee
 where $p(\Gamma,t)$ is the non-equilibrium probability of RNA being in a sub-state $\Gamma$ at time $t$. We note that rewriting \eqreff{eq:calF} gives
 \be\label{eq:p}
 p(\Gamma,t) = \frac{e^{-\beta G(\Gamma;\lambda_t)}}{e^{-\beta \mathcal{F}(\Gamma,t)}}.
 \ee
 Here the Boltzmann factor in the nominator is the same as the equilibrium distribution, \eqreff{eq:p_inf}. In the denominator appears local non-equilibrium free energy $\mathcal{F}$ of sub-state $\Gamma$ at time $t$ with Jarzynski's work as content, as in \eqreff{eq:JT}. Thus \eqreff{eq:p} represents that a sub-state $\Gamma$ at time $t$ endowed with energy through Jarzynski's work gets a higher probability of being realized, overcoming energy barriers due to the Boltzmann factor.

\section{Results}
In order to verify the local version of Jarzynski's work fluctuation theorem \eqreff{eq:JT}, we carry out Brownian motion simulations by solving the over-damped Langevin equation~\cite{sekimoto98,kubo}:
\be\label{eq:lan}
\zeta\dot{z}=-\nabla E(z;\lambda_t)+\xi,
\ee
which tells that a system's configuration is subject to a force in the direction of the lower energy during random walks caused by thermal fluctuations. Thermal fluctuation $\xi$ increases with temperature and should be uncorrelated at different times, i.e., $\left<\xi(t)\,\xi(t')\right>=2k_{B}T\zeta\delta(t-t')$. Here $\left<\cdot\right>$ indicates the average over all realizations and $\delta$ is the Dirac delta function. We set friction coefficient $\zeta=1$ and $k_{B}T=2$ and consider one-dimensional domain $0\le z\le L$ with $L=10$ with reflecting boundaries. 
We partition the domain into 20 bins, forming sub-states $\{\Gamma_i : i=1, 2, \cdots,20\}$. The initial probability distribution $p(z,0)$ is uniform on the domain. We consider two different external controls. The first one is single quenching, and the second is complex scheduling, including continuous changing, quenching, and relaxation.

\subsection{A quenched process}
At time $t=0$, we abruptly turn on bi-stable potential $E(z)$ depicted in Figure 1c and solve \eqreff{eq:lan}, taking 4k time steps with step size $\delta t=0.01$ for a single trajectory during $0\le t\le 40$. We repeat the experiment 200k times. The initial quenching would cause the initial equilibrium distribution into a non-equilibrium one and subsequently generate time series of non-equilibrium probability distributions $p(z,t)$. 

\begin{figure*}[t]
\centering
\includegraphics[width=16cm]{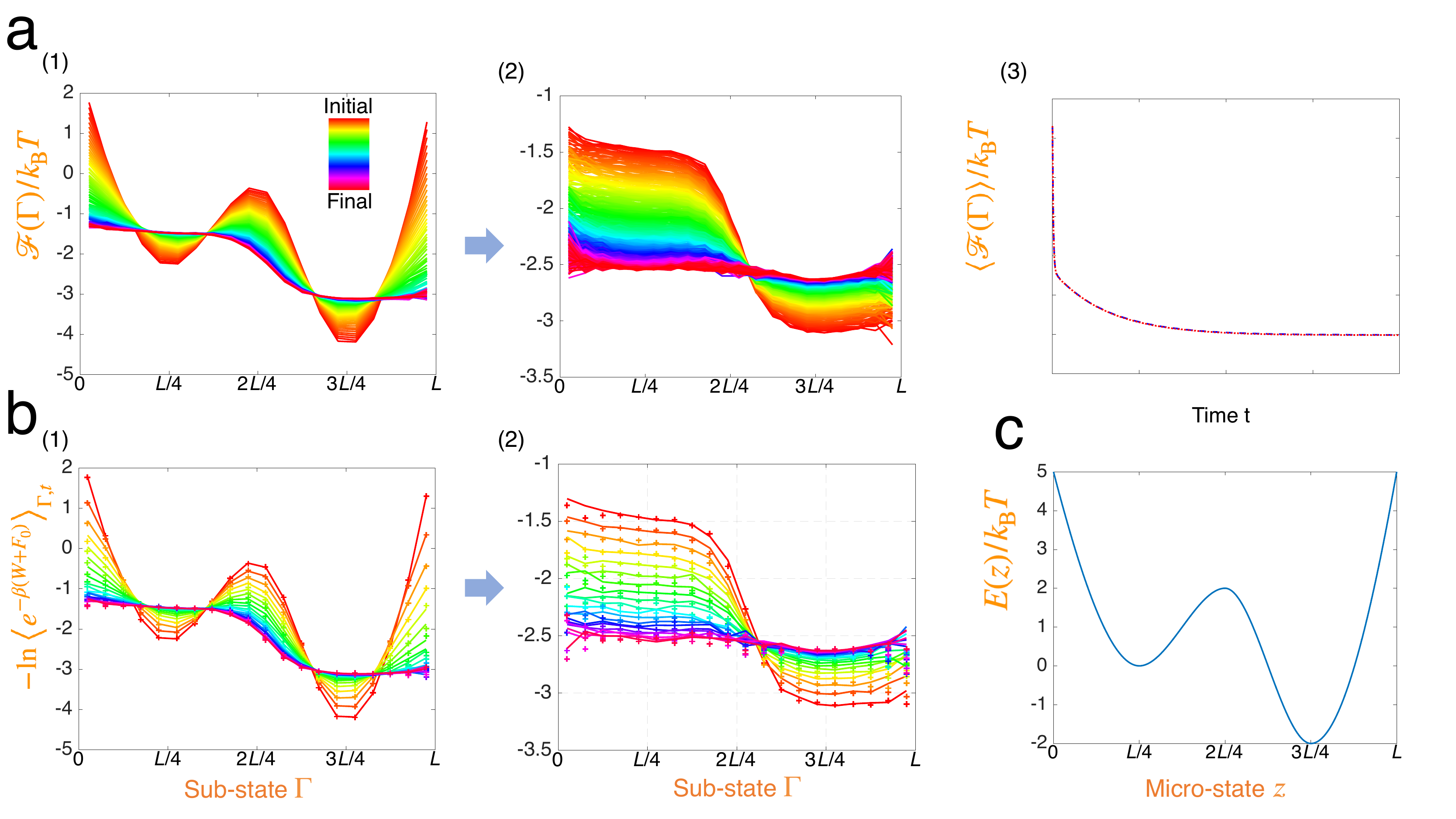}
\caption{Simulation Results for a Quenched Process:
(\textbf{a1}) The non-equilibrium free energy profiles $\mathcal{F}(\Gamma_i,t)$ for the first 50 steps ($0\le t\le 0.5$) are presented. Inset represents the color code for time steps. Panels a(2), b(1), and b(2) use the sam color code. 
(\textbf{a2}) The non-equilibrium free energy profiles $\mathcal{F}(\Gamma_i,t)$ for the remaining 3500 steps ($0.5< t\le 40$) are presented. 
(\textbf{b1}, \textbf{b2}) The color-coded plus symbols indicate (negative logarithm of) the ensemble averages of Jarzynski's work on the left-hand side of \eqreff{eq:JT} multiplied by $e^{-\beta F(\lambda_0)}$. Each ensemble comprises paths that reach $\Gamma_i$ at time $t$. The color-coded lines, the non-equilibrium free energy profiles, are drawn together for comparison. We show data for every five steps of $0\le t\le 0.5$ in (b1) and for every $200$ steps of $0.5<t\le40$ in (b2) to distinguish the symbols clearly.
(\textbf{a3}) Dashed blue line indicates the time series of the average over sub-states of non-equilibrium free energy, and dotted red line represents the time series of the average over sub-states of work-content that is represented in Panels (b1) and (b2).
(\textbf{c}) We show the energy profile turned on at time $0$.
}
\end{figure*}   

Panels (a1) and (a2) in Figure 1 show the time series of non-equilibrium free energy profiles. Initially, stochastic entropy $(-\ln p)$ is constant due to the uniformly distributed initial condition. Thus the conformational free energy $G$ in \eqreff{eq:G}, which resembles $E(z)$ depicted in Figure 1c in our case, determines the shape of non-equilibrium free energy $\mathcal{F}$ in \eqreff{eq:calF}. Since a particle is subject to a force in the direction of the lower energy, particles tend to move towards local minima, $z=L/4$ and $z=3L/4$ in the early rapid stage within 50 steps (see Panel (a1) in Figure 1). If stochastic entropy $(-\ln p)$ and conformational free energy $G$ are in balance, non-equilibrium free energy $\mathcal{F}$ becomes locally flat. Then the process of balancing the non-equilibrium free energy on a larger scale between regions $z<2L/4$ and $z>2L/4$ (see Panel (a2) in Figure 1). This large-scale process proceeds very slowly during the remaining 3.5k steps because a single particle that experiences energy $E(z)$ takes time in overcoming barrier region $L/4<z<2L/4$ in Figure 1c.

The color-coded plus symbols in Panels (b1) and (b2) of Figure 1 represent the ensemble average of Jarzynski's work on the left-hand side of \eqreff{eq:JT} multiplied by $e^{-\beta F(\lambda_0)}$ to match $\mathcal{F}$. Plus symbols are overlaid with the profiles of the non-equilibrium free energy for comparison. When quenching occurs, the particles gain different energies depending on their position. Since quenching occurs only once, gained work of individual particles is constant throughout the process. Therefore, it is worth questioning how the ensemble average of this work tends to dissipate. It is because the particles occupying a specific sub-state at any time come from different places and are mixed. If one considers a sub-state near $L/4$ in the early stage, particles mainly located in region $0\le z\le 2L/4$ would gather in that sub-state and are mixed so that the ensemble average of Jarzynski's work over those particles would be equalized. This mixing will proceed over time, similar to how the non-equilibrium free energies find balance.

In Panel (a3) in Figure 1, the dashed blue line represents the time series of the average over sub-states of the non-equilibrium free energy shown in Panels (a1) and (a2). The dotted red line shows the time series of the mean over sub-states of the work content represented in Panels (b1) and (b2). Both agree well and tend to dissipate.

 \subsection{A process including continuous changing, relaxation, and quenching}

 \begin{figure*}[t]
\centering
\includegraphics[width=16cm]{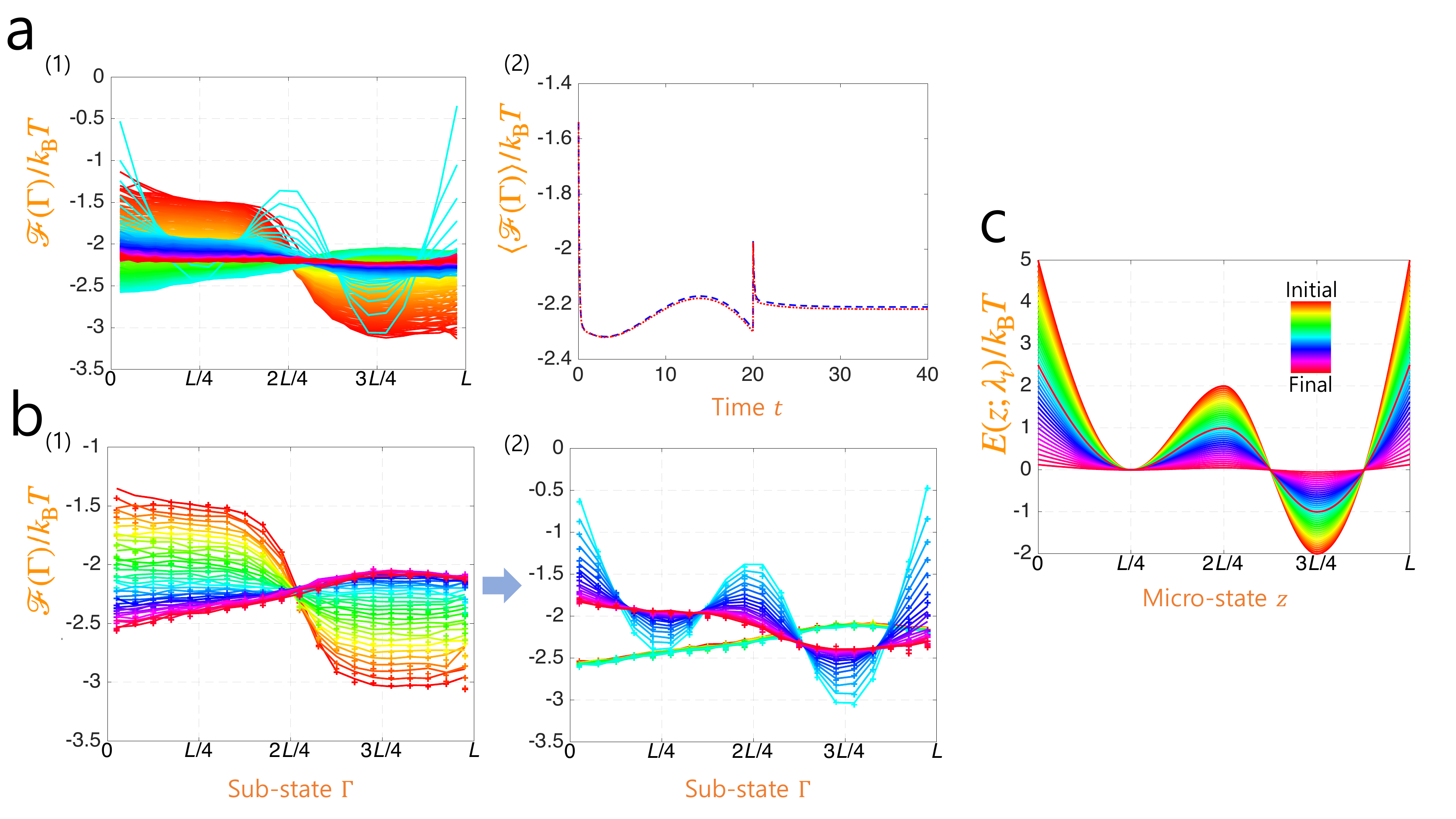}
\caption{Simulation results for a process composed of quenching, continuous changing, second quenching, and relaxation: 
(\textbf{a1}) The non-equilibrium free energy profiles $\mathcal{F}(\Gamma_i,t)$ during $0.5< t\le 40$ are presented.  
(\textbf{b1}, \textbf{b2}) The color-coded plus symbols indicate (negative logarithm of) the ensemble averages of Jarzynski's work on the left-hand side of \eqreff{eq:JT} multiplied by $e^{-\beta F(\lambda_0)}$. The color-coded lines, the non-equilibrium free energy profiles, are drawn together for comparison. We selected two time intervals that follow each other and exhibit patterns not seen in the first experiment. We show data for every 50 steps of $1< t\le 19$ in (b1) and for every five steps of $19<t\le 21$ in (b2). 
(\textbf{a2}) Dashed blue line indicates the time series of the average over sub-states of non-equilibrium free energy, and dotted red line represents the time series of the average over sub-states of work-content. 
(\textbf{c}) We show the energy profiles $E(z;\lambda_t)$ for $0\le t\le 40$. Inset represents the color code for time steps. Panels a(1), b(1), and b(2) use the sam color code.
}
\end{figure*}   

The scheduling of external control in the second process is as follows.
At time $t=0$, we abruptly turn on bi-stable potential $E(z)$ depicted in Figure 1c. Then gradually diminishes its amplitude by multiplying a time-dependent stiffness constant $\kappa(t)$ that varies from $1$ to $0$ for $0\le t<20$. At $t=20$, we apply the second quenching by setting $\kappa = 1/2$ and keep the value during the remaining steps so that the system would relax. Figure 2c represents the resulting $E(z;\lambda_t)=\kappa(t) E(z)$ for $0\le t\le 20$. We solve \eqreff{eq:lan} in the same way as in the first process, taking 4k time steps with step size $\delta t=0.01$ for a single trajectory during $0\le t\le 40$ and repeating the experiment 200k times. 

Figure 2a shows the time series of the non-equilibrium free energy profiles for $0.5<t \le 40$, omitting the initial relaxation stage that exhibits the same pattern as the case of the first process. In Panels (b1) and (b2) of Figure 2, the color-coded plus symbols (overlaid with $\mathcal{F}$ for comparison) represent the ensemble average of Jarzynski's work (on the left-hand side of \eqreff{eq:JT} multiplied by $e^{\beta F(\lambda_0)}$) for $1<t\le 21$. The selected interval includes the second quenching and thus exhibits patterns not seen in the first process. Data for final times in Panel (b1) of Figure 2 continues to the data for initial times in Panel (b2) of Figure 2. We see that $\mathcal{F}$ and work-content agree well during continuous changing, abrupt quenching, and the final relaxation. Panel (a2) of Figure 2 shows the averages over sub-states of $\mathcal{F}$ and work-content, and we see a convex pattern for $5<t<20$, which is worth mentioning. It is because the relaxation speed of particles during the large-scale relaxation stage cannot keep up with the change speed of the external control. Due to this, particles near $z=L/4$ lose energy, and particles near $z=3L/4$ gain energy but more quickly than the former, increasing the average $\mathcal{F}$ and balancing the two regions forcefully. Even after the forceful balance, the particles in the right-well keep gaining energy so that they have excess non-equilibrium free energy, which would be relaxed subsequently, decreasing the average $\mathcal{F}$ (see the final times of Panel (b1) of Figure 2).

\section{Application}

\subsection{Mechanism of RNA Translocation}
Here we analyze Ma and Schulten's simulation data for RNA-translocation by Rho hexameric helicase \cite{ma2015mechanism}.
Rho hexameric helicase\cite{thomsen2009}, a ring-shaped motor, translocates the substrate (RNA) during the rotary reaction from the state $R$ to $S$ \cite{ma2015mechanism}. The state $R$ represents a stable state taken from the crystal structure \cite{thomsen2009} with the ATP-mimics (ADP$\cdot$BeF3) at the six active sites being replaced by the equivalent ATP molecules (at four sites) and  ADP+Pi (at one site), leaving the remaining one site empty.
The state $S$ is identical to $R$ except 60$^\circ$ clockwise rotation. The transition $R\rightarrow S$ occurs
through subunit-subunit interface conformational changes, during which RNA is propelled. 

The transition $R\rightarrow S$ of the motor, which is not spontaneous, contains a dwell phase of a relative long duration and a motor-action phase \cite{liu2014mechanical}. During the dwell-phase, the motor is energetically charged through three step processes: (1) ATP binding in the empty site, (2) release of existing (say ``old'') ATP hydrolysis product (ADP+P$_i$), and (3) hydrolysis of ATP into ``new'' product (ADP+P$_i$) \cite{ma2015mechanism}. The subprocess (2) is considered rate limiting \cite{adelman2006mechanochemistry, chen2009adp}. Thus the subprocesses (1,3) take place at the very beginning of the transition,  
Rho is charged energetically, and waits until the subprocess (2) occurs. This state defines $I$ (see Figure 3a).
Then, subprocess (2) initiates the motor-action phase during which Rho translocates RNA through subunit-subunit 
interface conformational changes, which completes the cycle.

We interpret this cyclic process as a series of quenching, relaxation, and second quenching. In detail, we prepare the initial state as the state that subprocesses (1,3) have just taken place. Then, we take the first quenching that realizes subprocess (2), translocating RNA during relaxation. Then, we take the second quenching that realizes subprocess (1,3), which would turn the system into the initial state after another relaxation.
We do not involve this final relaxation and compare the initial probability distribution and the final non-equilibrium caused by the second quenching.

\subsection{Interpretation}
We introduce hypothetical external control $\lambda_t$ that turns on and off pre-determinedly the interaction energy between Rho hexameric helicase, ATP, and ADP+P$_i$. For an initial state, we set $\lambda_t$ (for $t<0$) as the state where the subprocesses (1,3) have just taken place from the state $R$. In detail, regarding subprocess (1), the interaction energy between ATP and an active site of Rho is turned on. For subprocess (2), the interaction energy between the existing ATP hydrolysis product (ADP+P$_i$) and Rho is kept turning on, describing the state where the ``old'' product is not released. Concerning subprocess (3), We change the interaction between an ATP and Rho to the interaction between ADP+P$_i$ and Rho, describing the hydrolysis of ATP into a ``new'' product (ADP+P$_i$). 
Let us shorten this initial state as $A$, i.e., $\lambda_t=A$ for $t<0$. 

Then, the equilibrium probability distribution $p_0$ of this state reads:
\be\label{eq:p0}
p_0(\Gamma_i; A) =\frac{e^{-\beta G(\Gamma_i; A)}}{e^{-\beta F(A) }},
\ee
where $G$ is the conformational free energy of sub-state $\Gamma_i$ at state $A$, and $F$ is the free energy at state $A$.
Ma and Schulten \cite{ma2015mechanism} defined reaction coordinates $\Gamma$ as a collection of positions of key residues at the six subunit-subunit interfaces, which contributes significantly to the relative motion of subunits. The transition $R\rightarrow S$ is then partitioned into 51 groups in terms of $\Gamma$ so that we have $\Gamma_i =  0, 1, \cdots, 50$, which we adopt as the sub-states. They calculated $G(\Gamma_i; A)-F(A)$, which we digitized to obtain $p_0(\Gamma_i;  A)$. We normalized so that $\sum_0^{50} p_0(\Gamma_i;A) = 1$. 
In Figures 3b and 3c, blue curves represent the conformation free energy and the initial probability distribution given $\lambda_t=A$.

At time $t=0$, we abruptly turn off the interaction energy concerning subprocess (2) so that the ``old'' ATP hydrolysis product (ADP+P$_i$) is released. Let us denote this quenched state by $B$ so that $\lambda_t=B$ for $0\le t< \tau$.
This quenching would initiate a motor-action phase, yielding the time-series of non-equilibrium probability distributions $p$:
\be
p(\Gamma_i, t) =  \frac{e^{-\beta G(\Gamma_i; B)}}{e^{-\beta \mathcal{F}(\Gamma_i,t) }},
\ee
where $\mathcal{F}(\Gamma_i,t)$ in the denominator is the non-equilibrium free energy of sub-state $\Gamma_i$ at time $t$. 
We do not change $\lambda_t$ until Rho translocates RNA during relaxation. Let $\tau$ be the time long enough to  complete the process. We note that $\tau$ should be fixed during the repeat of this (hypothetical) experiment. After the relaxation, the final equilibrium probability distribution $p_1$ would be:
\be
p_1(\Gamma_i; B) =\frac{e^{-\beta G(\Gamma_i; B)}}{e^{-\beta F( B) }}.
\ee
Ma and Schulten \cite{ma2015mechanism} calculated $G(\Gamma_i; B)-F(B)$, which we digitized to obtain $p_1(\Gamma_i; B)$. We normalized so that $\sum_0^{50} p_1(\Gamma_i;B) = 1$. 
In Figures 3b and 3c, orange curves represent the conformation free energy and the final probability distribution given $\lambda_t=B$.
 At time $t=\tau$, we perform the second quenching by setting $\lambda_\tau=A$, the same state as the initial one. Then the final equilibrium distribution turns into a non-equilibrium one: 
\be\label{eq:p1}
p(\Gamma_i, \tau) = \frac{e^{-\beta G(\Gamma_i;A)}}{e^{-\beta \mathcal{F}(\Gamma_i, \tau) }}.
\ee
We apply the local work fluctuation theorem to this process $\lambda_t$ for $0\le t\le \tau$ to obtain
\be\label{eq:JT2}
\left<e^{-\beta W}\right>_{\Gamma_i, \tau}=e^{-\beta \left[\mathcal{F}(\Gamma_i,\tau)-F(A)\right]}.
\ee
We can obtain the right-hand side of \eqreff{eq:JT2} by comparing the initial probability distribution \eqreff{eq:p0} and the final non-equilibrium probability distribution \eqreff{eq:p1} as follows:
\be
\ln\left[\frac{p(\Gamma_i,\tau)}{p_0(\Gamma_i;A)} \right] = \beta \left[\mathcal{F}(\Gamma_i, \tau) - F(A)\right],
\ee
which is represented in Figure 3d.

\subsection{Analysis}
 \begin{figure*}[t]
\centering
\includegraphics[width=16cm]{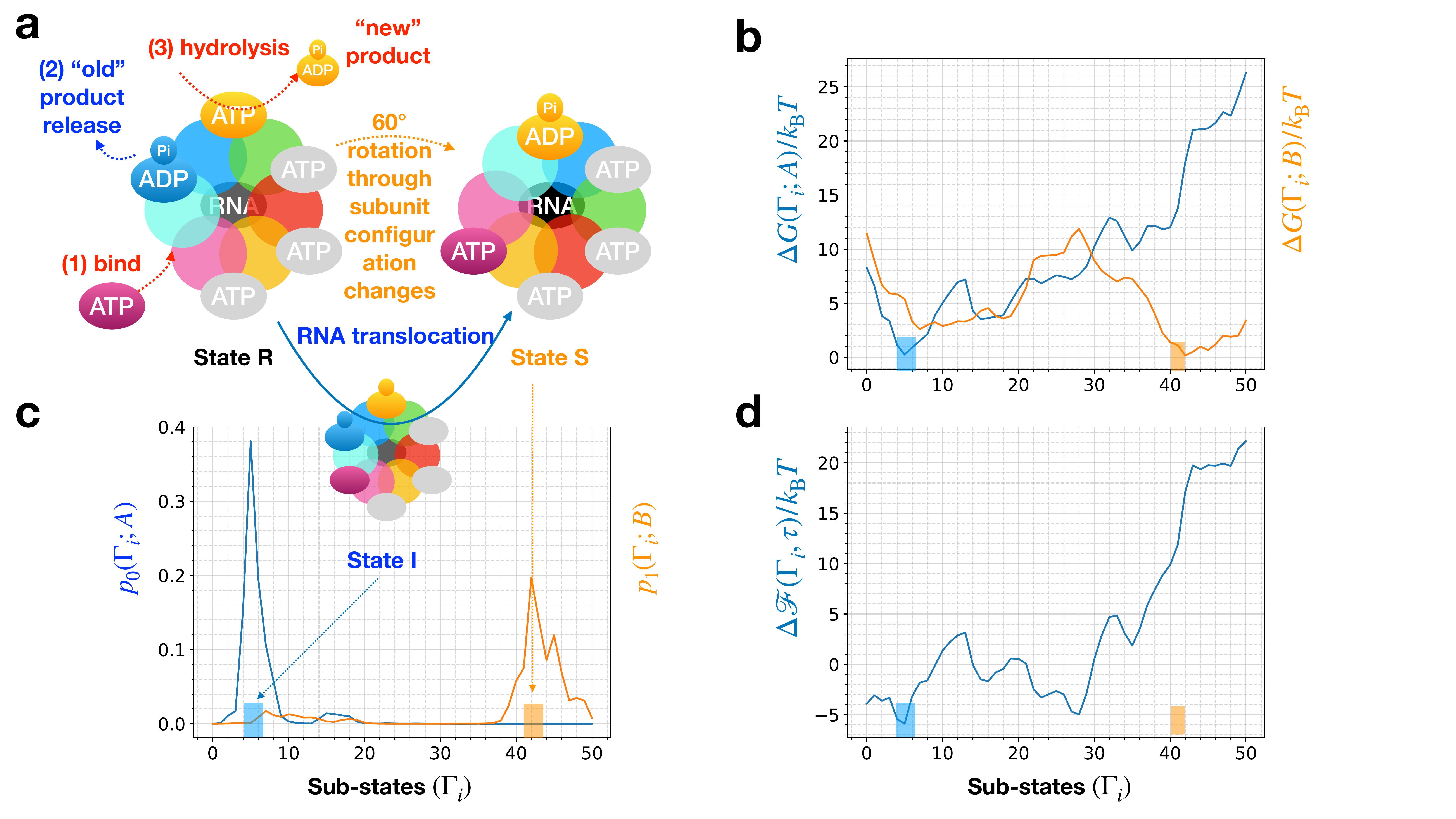}
\caption{Analysis of the simulation result for RNA translocation by Rho hexameric helicase:
(\textbf{a}) The state $R$ represents a stable state taken from the crystal structure \cite{thomsen2009} with the ATP-mimics at the six active sites being replaced by the equivalent ATP molecules (at four sites) and  ADP+Pi (at one site), leaving the remaining one site empty. The state $S$ is identical to $R$ except for 60$^{\circ}$ clockwise rotation of subunit-subunit interface conformations (which are schematically represented by different colors), which is achieved by conformational changes of subunit-subunit interfaces along the reaction coordinate. This rotary reaction $R\rightarrow S$ is energetically prepared during a dwell phase: (1) ATP binding at the empty site and (2) ATP hydrolysis in another already occupied (by ATP) site \cite{ma2015mechanism}. This state defines state $I$.
The motor reaction phase is then initiated by the release of ATP-bound product (ADP+P$_i$) so that the stored energy during the dwell phase causes the rotary reaction.
(\textbf{b}) Conformational free energies $G(\Gamma_i; A)-F(A)$ in blue and $G(\Gamma_i; B)-F(B)$ in orange are presented.
(\textbf{c}) Initial probability distribution $p_0(\Gamma_i; A)$ in blue and final probability distribution $p_1(\Gamma_i; B)$ in orange are presented. The most populated sub-state in $p_0$ is state $I$, and that in $p_1$ is state $S$.  
(\textbf{d}) The non-equilibrium free energy profile $\mathcal{F}(\Gamma_i, \tau)-F(A)$ is presented.
}
\end{figure*}   

Let us follow what happens to the ensemble of rho hexameric helicases (with RNAs) during external control $\lambda_t$ varies from $0$ to $\tau$.  
The initial probability distribution with $\lambda_t=A$ (blue curve in Figure 3c) tells that most instances would be near $\Gamma_i=5$. If the molecules' ``old'' ATP hydrolysis products are released, or $\lambda_t$ is set to $B$, the conformational free energy profile would change from the blue curve in Figure 3b to the orange one so that the ensemble of conformations near $\Gamma_i=5$ would have excess non-equilibrium free energy. As time flows, the excess free energy would become flat locally, covering regions $0 < \Gamma_i < 28$. Subsequently, global balancing would proceed very slowly between regions $\Gamma_i<28$ and $\Gamma_i>28$. This stage corresponds to individual molecules' overcoming the free energy barrier near $\Gamma_i=28$. As the time approaches $\tau$, the non-equilibrium free energy would become flat for all $\Gamma_i$, and most instances would be near $\Gamma_i=41$ as the orange curve in Figure 3c indicates.
Then, we set $\lambda_t=A$; then again, they would have excess free energy, which is shown in Figure 3d. 

According to the local work fluctuation theorem \eqreff{eq:JT2}, the excess free energy also tells that 
how much the molecules have been energized during the process, especially by the two quenching steps.
We note that $\lambda_t=A$ is the state where an ATP is already hydrolyzed before the process begins at time $t=0$. During the intermediate $0<t<\tau$, the release of the ``old'' ATP product causes a motor action. Only at time $t=\tau$, when we set $\lambda_\tau$ to $A$, the hydrolysis of new ATP happens. So the net effect of the two quenching upon a single instance of the system is the hydrolysis of one ATP molecule.

At time $\tau$, it is the most probable that the system is in state $\Gamma_i\approx 41$ and its non-equilibrium free energy value, $\mathcal{F}(\Gamma_i=41,\tau)-F(A)$, is approximately $12 k_{\rm B}T$. Provided that ATP hydrolysis energy is $20 \sim 25 k_{\rm B}T$, depending on the environmental condition \cite{ROSING1972275}, the efficiency of the motor action corresponds to $48 \sim 60\%$. It is interesting that the non-equilibrium free energy contains work values for another states at time $\tau$. For example, sub-state $\Gamma_i=45$,  the value of $\Delta\mathcal{F}(\Gamma_i=45,\tau)$ is about $20 k_{\rm B}T$, corresponding to $80 \sim 100\%$ efficiency. The final probability value for this state is $p_1(\Gamma_i=45,\tau)=0.12$, indicating that $12\%$ of the molecules in the ensemble are achieving this efficiency. We note that some literature reports near $100\%$ efficiency of molecular motors~\cite{effi}.
   

\section{Conclusions}
We have verified the local version of the work fluctuation theorem linking Jarzynski's work and local non-equilibrium free energy. In various conditions of external controls, including quenching, relaxation, and continuous changing, the non-equilibrium free energy and Jarzynski's work agree very well for each substate at any time. As an application, we analyzed a simulation result of RNA translocation by Rho hexameric helicase. By treating one cycle of a rotary reaction mechanism as a series of quenching, relaxation, and second quenching, we could obtain Jarynski's work content of each substate of Rho at the end of the translocation process. This theory would enable one to extract non-equilibrium work content from modern single-molecule experiments and simulations.

\section{Appendix}
\subsection{Proof of \Eqreff{eq:JT}}
We consider the local non-equilibrium free energy $f$ for micro-state $z$:
\be\label{eq:f}
 f(z,t) = E(z;\lambda_t)+\frac{1}{\beta}\ln p(z,t),
\ee
Let $\{\Gamma_i\}$ be sub-states. We calculate the following average of local non-equilibrium free energy for microstates $z\in\Gamma_i$:
\begin{eqnarray}\nonumber
\left<e^{-\beta f(z,t)}\right>_{z\in\Gamma_i}&=&\int_{z\in\Gamma_i}e^{-\beta f(z;\lambda_t)}\frac{p(z,t)}{p(\Gamma_i,t)}\,dz\\
\nonumber
&=&\int_{z\in\Gamma_i}\frac{e^{-\beta E(z;\lambda_t)}}{p(\Gamma_i,t)}\,dz=\frac{e^{-\beta G(\Gamma_i;\lambda)}}{p(\Gamma_i,t)}\\
\label{eq:fluct}
&=& e^{-\beta \mathcal{F}(\Gamma_i,t)},
\end{eqnarray}
where we used \eqreff{eq:G}, \eqreff{eq:calF}, and \eqreff{eq:f}.
We will use this fluctuation theorem linking $f$ and $\mathcal{F}$ to prove the local work fluctuation theorem.
 
Let us consider external control $\lambda_t$ from $0$ to $\tau$. We assume that the initial probability for the reversed process is the same as the final probability of the forward process, i.e., $p'(z'_0,0)=p(z,\tau)$. By the property of conditional probability, we have 
${p\left[\{z_t\}_{0\le t\le \tau}\right]}={p\left[\{z_t\}_{0\le t\le \tau}|z_0\right]}p(z_0,0)$ and similarly,
$p'\left[\{z'_t\}_{0\le t\le\tau}\right]=p'\left[\{z'_t\}_{0\le t\le\tau}|z'_0\right]p'(z'_0,0)$.
Using the microscopic reversibility condition \eqreff{eq:rev} and \eqreff{eq:f} give the following:
\begin{eqnarray}\nonumber
\frac{{p\left[\{z_t\}_{0\le t\le \tau}\right]}}{p'\left[\{z'_t\}_{0\le t\le\tau}\right]} &=& e^{-\beta Q +\beta \Delta E -\beta\Delta f}\\
\label{eq:w_f}
&=&e^{W -\Delta f},
\end{eqnarray}
where $\Delta E=E(z_\tau,\tau)-E(z_0,0)$ and $\Delta f=f(z_\tau,\tau)-f(z_0,0)$, and we used the first-law of thermodynamics \eqreff{eq:1st-law}.
Assuming the initial state is equilibrium, i.e., $f(z_0,0)=F(\lambda_0)$,
we prove \eqreff{eq:JT}:
\begin{eqnarray*}
\left<e^{-\beta W}\right>_{\Gamma,\tau} &=& \int e^{-\beta W}
\frac{p\left[\{z_t\}_{0\le t\le \tau}\right]}{p(\Gamma_i,\tau)}\, d\left[\{z_t\}_{0\le t\le \tau}\right]\\
 &=&\int e^{-\beta\Delta f}\frac{p'\left[\{z'_t\}_{0\le t\le\tau}\right]}{p(\Gamma_i,\tau)}\,d\left[\{z_t\}_{0\le t\le \tau}\right]\\
 &=& e^{\beta F(\lambda_0)}\int_{z_{\tau}\in\Gamma_i}e^{-\beta f(z_\tau,\tau)}\frac{p(z_\tau,\tau)}{p(\Gamma_i,\tau)}\,dz_\tau\\
 &=& e^{-\beta \left[\mathcal{F}(\Gamma,\tau)-F(\lambda_0)\right]},
\end{eqnarray*}
 where we used the fluctuation theorem that links $f$ and $\mathcal{F}$ in \eqreff{eq:fluct}, $p'(z'_0,0)=p(z_\tau,\tau)$, \eqreff{eq:w_f}, $\int_{z_\tau\in\Gamma_i} e^{-\beta f(z_\tau,\tau) p'\left[\{z'_t\}_{0\le t\le\tau}\right] }d\left[\{z'_t\}_{0\le t\le \tau}\right]=\int_{z_\tau\in\Gamma_i}e^{-\beta f(z_\tau,\tau)}p(z_\tau,\tau)\,dz_\tau$
 and $d\left[\{z_t\}_{0\le t\le \tau}\right]=d\left[\{z'_t\}_{0\le t\le \tau}\right]$ due to the time-reversal symmetry.

\noindent
\section*{Acknowledgments}
The author was supported by the National Research Foundation of Korea Grant funded by the Korean Government (NRF-2016R1D1A1B02011106), and in part by Kwangwoon University Research Grant in 2019.

%

\end{document}